\documentclass{JHEP3}
\usepackage{epsfig,amsmath,amsthm,latexsym}
\newcommand{\be}{\begin{equation}}
\newcommand{\ee}{\end{equation}}
\newcommand{\bea}{\begin{eqnarray}}
\newcommand{\eea}{\end{eqnarray}}
\newcommand{\IR}{\mathbb{R}} 

\def\IZ{\relax\ifmmode\hbox{Z\kern-.4em Z}\else{Z\kern-.4em Z}\fi}
\newcommand{\IS}{{\bf S}}

\newcommand{\non}{\nonumber \\}

\def\half{{1 \over 2}} 

\def\del{{\partial}}
\def\room{~\rule[-2mm]{0mm}{8mm}}

 \def\cdr{{d \over d\rho_R}\,}

\def\wV{\widetilde{V}}

\def\al{\alpha}

\def\hal{{\hat \alpha}}

\newcommand{\sbsection}[1]{\vspace{.5cm} \noindent {\bf #1}}
\def\room{~\rule[-2mm]{0mm}{8mm}}
\def\presub{\vspace{.5cm} \noindent}

\def\bi{\begin{itemize}} \def\ei{\end{itemize}}

\def\({\left(} \def\){\right)}
\def\[{\left[} \def\]{\right]}
\preprint{{\tt hep-th/0607129}}

\title{ \center{Analytic Evidence for Continuous Self Similarity of the Critical Merger Solution}}

\author{
Vadim Asnin\footnotemark[1], Barak Kol\footnotemark[2]
 and Michael Smolkin\footnotemark[3] \\
 Racah Institute of Physics\\
 Hebrew University \\
 Jerusalem 91904,
 Israel\\

 \footnotemark[1] {\tt vadima@pob.huji.ac.il} \\
 \footnotemark[2] {\tt barak\_kol@phys.huji.ac.il} \\
 \footnotemark[3] {\tt smolkinm@phys.huji.ac.il} }

\abstract{The double cone, a cone over a product of a pair of
spheres, is known to play a role in the black-hole black-string
phase diagram, and like all cones it is continuously self similar
(CSS). Its zero modes spectrum (in a certain sector) is determined
in detail, and it implies that the double cone is a co-dimension 1
attractor in the space of those perturbations which are smooth at
the tip. This is interpreted as strong evidence for the double
cone being the critical merger solution. For the
non-symmetry-breaking perturbations we proceed to perform a fully
non-linear analysis of the dynamical system.
 The scaling symmetry is used to reduce the dynamical system from a 3d
phase space to 2d, and obtain the qualitative form of the phase
space, including a non-perturbative confirmation of the existence
of the ``smoothed cone''.}

\begin{document}

\section{Introduction and Summary}





\begin{figure}[t!]
\centering \noindent
\includegraphics[width=7cm]{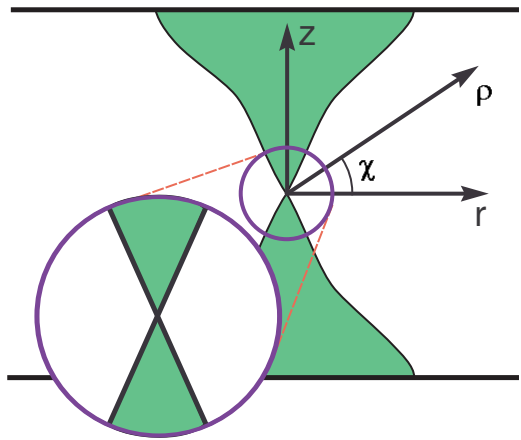}
\caption[]{The merger metric. $r$ is the radial coordinate in the
extended directions, $z$ is periodically compactified, while time
and angular coordinates are suppressed. The heavy lines denote the
horizon of a static black object which is at threshold between
being a black-hole and being a black-string. The naked singularity
is at the $\times$-shaped pinching (horizon crossing) point. Upon
zooming onto the encircled singularity it is convenient to replace
$(r,z)$ by radial coordinates $(\rho,\chi)$ radial coordinates. We
shall be mostly interested in the ``critical merger solution'' --
the local metric near the singularity, namely the encircled
portion of the metric (in the limit that the circle's size is
infinitesimal).} \label{merger-figure}
\end{figure}

The phase diagram of the black-hole black-string transition (see the
reviews \cite{review,HOrev}) was conjectured in \cite{TopChange} to
include a ``merger'' point -- a static vacuum metric (see figure
\ref{merger-figure}) which lies on the boundary between the
black-string and black-hole branches. It can be thought as either a
black string whose waist has become so thin that it has marginally
pinched, or as a black-hole which has become so large that its poles
had marginally intersected and merged (and hence the name
``merger''). The metric cannot be completely smooth as it
interpolates between two different space-time topologies, but it may
have only one naked singularity.

The black-hole black-string system has been the subject of intensive
numerical research
\cite{Wiseman1,KPS2,KudohWiseman1,KudohWiseman2,KKR}. Naturally, the
merger space-time itself is unattainable numerically since it
includes a singularity, but it may be approached by following either
of the two branches far enough. Indeed, all the available data
indicates that the black-string and the black-hole branches approach
each other, in accord with the merger prediction.

At merger the curvature is unbounded around the pinch point. One
defines the ``critical merger solution'' to be the local metric
around the pinch point (at merger), namely the one achieved
through a zooming limit around the point. It is natural to predict
\cite{TopChange,BKscaling} that the critical merger solution will
lose all memory of the macroscopical scales of the problem (the
size of the extra dimension and the size of the black hole) and
moreover be self-similar, namely invariant under a scaling
transformation. The central motivation of this paper is \emph{to
determine the critical merger solution}.

Self-similar metrics belong to one of two classes: Continuous
Self-Similarity (CSS) or Discrete Self-Similarity (DSS): while a CSS
metric is invariant under any scale transformation and can be
pictured as a cone, a DSS metric is invariant only under a specific
scaling transformation (and its powers) and can be pictured as a
wiggly cone with its wiggles being log-periodic (see figure
\ref{cone-illus}). A key question is: \emph{Is the critical merger
solution CSS or DSS}?
 \EPSFIGURE{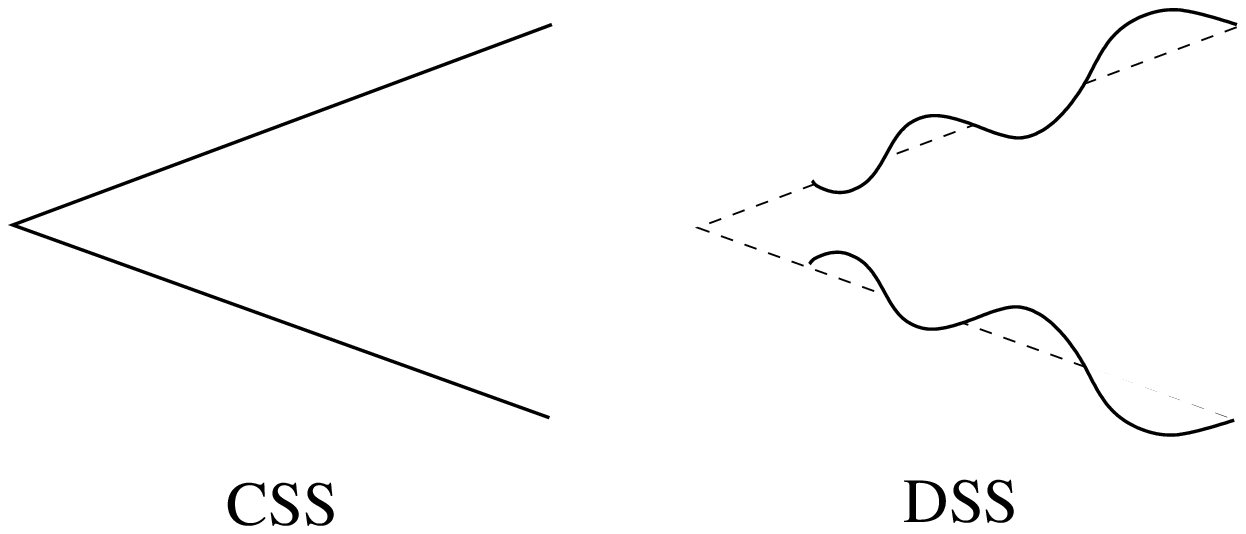,width=7cm}{An illustration of
a continuously self-similar geometry (CSS) as a cone, and a
discretely self-similar geometry (DSS) as a wiggly cone. The
singularity is at the tip. \label{cone-illus}}

The significance of the critical solution is that critical exponents
of the system near the merger point are determined by properties of
this solution (this is known to be the case in the closely related
system \cite{BKscaling} of Choptuik critical collapse
\cite{Choptuik,GundlachRev}).

\presub {\bf Considerations and plan}. The direct way to settle
our question would be through numerical simulation of the system:
one would need to obtain solutions which are ever closer to the
merger and with ever higher resolution near the high curvature
region (where the singularity is about to form). This computation
would be comparable in difficulty to Choptuik's original discovery
of critical collapse \cite{Choptuik} in that it requires
successive mesh refinements over several orders of magnitude.

However, since we are asking a local question, we may expect or
hope that a local analysis would suffice, namely the analysis of
the metric close to the singularity. On the one hand a local
analysis is disadvantaged relative to a solution of the whole
system in being indirect and therefore its results need some
interpretation which reduces certainty. On the other hand, a local
analysis supplies more insight into the mechanism that determines
the local metric, and it is easier to perform. These latter
advantages induced us to prefer the local analysis for the current
study. The demanding full (numerical) analysis is yet to be
performed (see however the suggestive but inconclusive results in
\cite{KolWiseman}).

Our first assumption is that the local metric is self-similar, as
discussed above. Our second working assumption is that if several
self-similar metrics exist then the one actually realized by the
system is the metric which is most attractive, or most stable, in an
appropriately defined manner.

One local self-similar solution, the ``double-cone'' has been
known for a while \cite{TopChange}. In terms of the $(\rho,\chi)$
coordinates defined in figure \ref{merger-figure} it is given by
\be
 ds^2= d\rho^2 + {1 \over D-2}\, \rho^2 \[ d\chi^2+\cos^2(\chi)\, dt^2 + (D-4)\,
 d\Omega^2_{D-3} \] \label{d-cone} ~,\ee
 where $t$ denotes Euclidean time (since the solutions are static we
may work either with a Lorentzian or with a Euclidean signature),
and $d\Omega^2_{D-3}$ is the standard metric on the $\IS^{D-3}$
sphere.
Let us recall some of its properties. The $(\chi,t)$ portion of
the metric is (conformal to) the two-sphere $\IS^2$. Thus the
metric is a cone over a product of spheres $\IS^2 \times
\IS^{D-3}$ which is the origin of its name. Its isometry group is
$SO(3)_{\chi,t} \times SO(D-2)_\Omega$, which is an enhancement
relative to the generic $SO(2)_t \times SO(D-2)_\Omega$ isometry
of the system. The double cone is smooth everywhere except for the
tip $\rho=0$. It is manifestly CSS under the transformations $\rho
\to e^\alpha\, \rho$ for any $\alpha$. Finally, a linear analysis
of perturbations around the double cone preserving the full $SO(3)
\times SO(D-2)$ isometry reveals \cite{TopChange} an oscillating
nature (as a function of $\rho$) for low enough dimensions,
$D<10$.

The current research started from a confusion regarding the CSS/DSS
nature of the critical merger solution. \cite{TopChange} assumed CSS
for simplicity and found the double cone. \cite{KolWiseman} found
good but not overwhelming evidence for the double cone in numerical
solutions. In \cite{BKscaling} a close relation between the merger
and critical collapse was discovered,\footnote{See also
\cite{SorkinOren} which followed and studied the dimensional
dependence of the Choptuik scaling constants.} raising the
possibility that just as the critical collapse solution is DSS, the
critical merger could be DSS as well. Moreover, the linearized
oscillations, mentioned in the preceding paragraph were realized to
be analogous to GHP oscillations (Gundlach-Hod-Piran
\cite{Gundlach96,HodPiran}) in critical collapse. In critical
collapse the log-period of GHP oscillations is known to be
essentially the log-period of the DSS critical solution. Therefore
\cite{BKscaling} viewed the oscillations as pointing towards a DSS
nature.


Accordingly, our first objective was to find a DSS solution to the
system. Since we saw little hope in finding an analytic solution, we
turned to a numerical method. Rather than simulate the whole system
and tune a parameter for criticality, we followed \cite{Gundlach95}
and imposed periodic boundary conditions.
 We used two different algorithms. The first was of a relaxation
type: the 3 fields are solved for iteratively using a selected 3 out
of the 5 equations. This method involves an interesting interplay
between the usual local variables (the fields) and certain global
variables. The essential idea in the second algorithm is to take as
a merit function the sum of squares of all 5 equations. Altogether,
despite considerable work the code never converged to a (new) DSS
solution, but rather to the double-cone. Therefore we relegate the
description of algorithm and implementation to appendix
\ref{search-section} and choose to detail only the second approach.

The apparent paradox between the existence of GHP oscillations ad
the absence of a DSS solution is explained, in hindsight, by the
fact that while DSS indeed implies GHP oscillations the converse is
incorrect: oscillations may arise also around CSS solutions.

The misguided fruitless search for a DSS solution pointed us
towards a different effort, which is the focus of this paper: one
could \emph{analyze the linear stability of the double cone}. If
it is found to be unstable (in a sense to be described below) then
it is very unlikely to be the metric chosen by the black-hole
black-string system.

Indeed, the asymptotic boundary conditions, including the compact
nature of the extra dimension, can be viewed as an asymptotic
perturbation of the critical merger solution. By assumption, this
perturbation is irrelevant near the singularity. In addition one
could consider turning on various perturbations far away from the
system, such as putting the black object in a non-flat but
low-curvature background or turning on a cosmological constant
(note that these perturbations belong to a wider class -- the
first does not obey the generic isometries and the second perturbs
the equations). If the double-cone is found to be unstable to some
asymptotic perturbation it would be unlikely to be realized as the
critical merger solution. On the other hand, if it is found to be
stable that would make it a viable candidate for being the
critical merger solution.

\presub {\bf Stability}. The preceding discussion motivates us to
formulate our stability criterion. Actually, we are seeking a
solution which is not absolutely stable to asymptotic
perturbations, but one which has a single unstable such mode --
this is the mode which corresponds to motion on the branch of
black-string (or black-hole) solutions away from merger.\footnote{
More precisely, motion onto the two branches, that of a black-hole
and that of a black-string, is associated with two modes defined
up to multiplication by $\IR_+$ and these modes are not
necessarily related by a multiplication by -1. See the last
paragraph of section 3 for further details.}
Therefore we define a self-similar solution to be \emph{stable}
\emph{if all but one asymptotic perturbation are irrelevant at the
singularity}, and we proceed to define ``irrelevant'' and
``asymptotic perturbation''.

Each mode can be characterized by its $\rho$ (see figure
\ref{merger-figure}) dependence, which must be a power law, since
the background is continuously self-similar and therefore the
scaling generator can be diagonalized simultaneously with the small
perturbations operator (the Lichnerowicz operator) -- this will be
seen explicitly in section \ref{pert-section}. For each mode we
define a constant $s$ through \be
 \delta g^{\mu}_{\nu} \sim \rho^s ~, \label{def-s} \ee
where $\delta g_{\mu \nu}$ is the perturbation to the metric, and in
general $s$ could be complex. Actually since the modes are
determined by a system of second order ordinary differential
equations (ODEs), the modes come in pairs, and we denote the
corresponding pair of $s$ constants by $s_\pm$, ordered such that
$\Re(s_-) \le \Re(s_+)$.

We refer to a mode as \emph{irrelevant} if it is negligible close to
the tip (the singularity), namely if \be
 s > 0 ~. \label{irrelevant} \ee

We define the \emph{asymptotic perturbations} as those
corresponding to $s_+$. The rational behind the definition is
common-place: for example in electro-statics, solutions of the 3d
Laplace equations come with two possibilities for the radial
dependence for each angular number $l$, being either $r^l$ or
$r^{-l-1}$. The $r^l$ mode is interpreted as an asymptotic
perturbation, while the $r^{-l-1}$ is interpreted as a
perturbation to the source which lies at the origin. Our
definition is ambiguous when $\Re(s_1) = \Re(s_2)$, but it will
happen only for the $l=0$ case which is studied in detail in
section \ref{non-pert-section}, and will turn out to pose no
problem. \footnote{While our definition is intuitively clear it
differs from the usual definitions requiring smoothness and
normalizability at the tip. Since the double cone is singular the
usual prescriptions do not apply. Conceptually one could determine
the perturbation spectrum around the smoothed cone demanding the
standard smoothness and normalizability at the smooth tip, and
then rescale towards the double-cone (see figure
\ref{ScaledCones}). It would be interesting to test whether this
limit would result in our ``$s_+$ prescription''. Another
possibility would be to perform a non-linear admissability
analysis of the modes, which we indeed perform in the $l=0$
sector, as explained below.}

Altogether our definition of stability as the case when all
asymptotic perturbations but one are irrelevant at the tip means
that \be
 \fbox{$~\room
 s_+ > 0
 ~~$} \label{stability} \ee
 for all but one perturbation. It can be said that such a solution
is a co-dimension 1 attractor for asymptotic perturbations. We note
that this definition differs from other common definitions of
stability which involve the positivity of the Lichnerowicz operator
or absence of imaginary frequencies in modes with time dependence.


\presub {\bf Method and results}. In section \ref{pert-section} we
determine the spectrum of $s$ constants (\ref{def-s}) for all
perturbations with the generic $SO(2) \times SO(D-2)$ isometry. We
use an action approach: we write down the most general ansatz
consistent with the isometry which includes 5 fields which depend
on two variables, compute the action and expand it to second order
in perturbations. Then we fix a gauge, derive the corresponding
pair of constraints from the action and the (other) 3 equations of
motion. The latter are solved by separation of variables and the
solutions are then tested against the constraints. Our result for
the spectrum is summarized in (\ref{spectrum}). It consists of two
families of solutions $s^s_\pm(l),\, s^t_\pm(l)$, a scalar and a
tensor with respect to $\IS^2_{\chi,t}$.

For all but one\footnotemark[2] $l=0$ mode we find that $s_+> 0$
and hence as explained above the double cone is a viable candidate
to be the critical merger solution. Combining this with the fact
that even after our search for DSS solutions described in appendix
\ref{search-section} the double-cone remains the only known
self-similar solution with these symmetries, \emph{we interpret
the result as strong evidence that the double cone is indeed the
critical merger solution}.

\presub {\bf Non-linear spherical perturbations.} The spherical
($l=0$) perturbations are special. There are two such modes and in
\cite{TopChange} evidence was given that there are two specific
linear combinations that generates ``smoothed cones'' (see figure
\ref{ScaledCones}) by attempting a Taylor expansion of the fields
and equations around the smooth tip of an assumed smoothed cone
and finding no obstruction to a solution. In section
\ref{non-pert-section} we confirm this by an analysis of the
qualitative features of the full non-linear dynamical system. The
perturbation associated with the smoothed cones is precisely the
special relevant mode mentioned above that moves the solution off
criticality and along the solution branch\footnotemark[2]. Any
other linear combination is seen to be highly singular at the tip,
which justifies us in discarding it (and it is consistent with our
``$s_+$ prescription'').

From the analytical point of view the qualitative analysis shows
an interesting feature. The system is non-integrable (see the
discussion below eq. \ref{K2}).  The qualitative dynamics of 2d
phase spaces is quite limited and never chaotic while in higher
dimensions chaos is common. The phase space of this system is 3d:
there are two phase space dimensions for each of the two modes
minus a constraint. However, the dynamical system inherits the
scaling symmetry of the background. One can define a reduced 2d
phase space system by choosing a plane transversal to the symmetry
flow, and then define a reduced dynamical flow to be the original
flow projected onto the plane through the symmetry flow. This 2d
phase space system is now amenable to analysis through the
determination of equilibrium points: focal, nodal or saddle, and
we are able to solve for the full qualitative features. The
solution is summarized in figure \ref{phase-space5d}.
 \EPSFIGURE{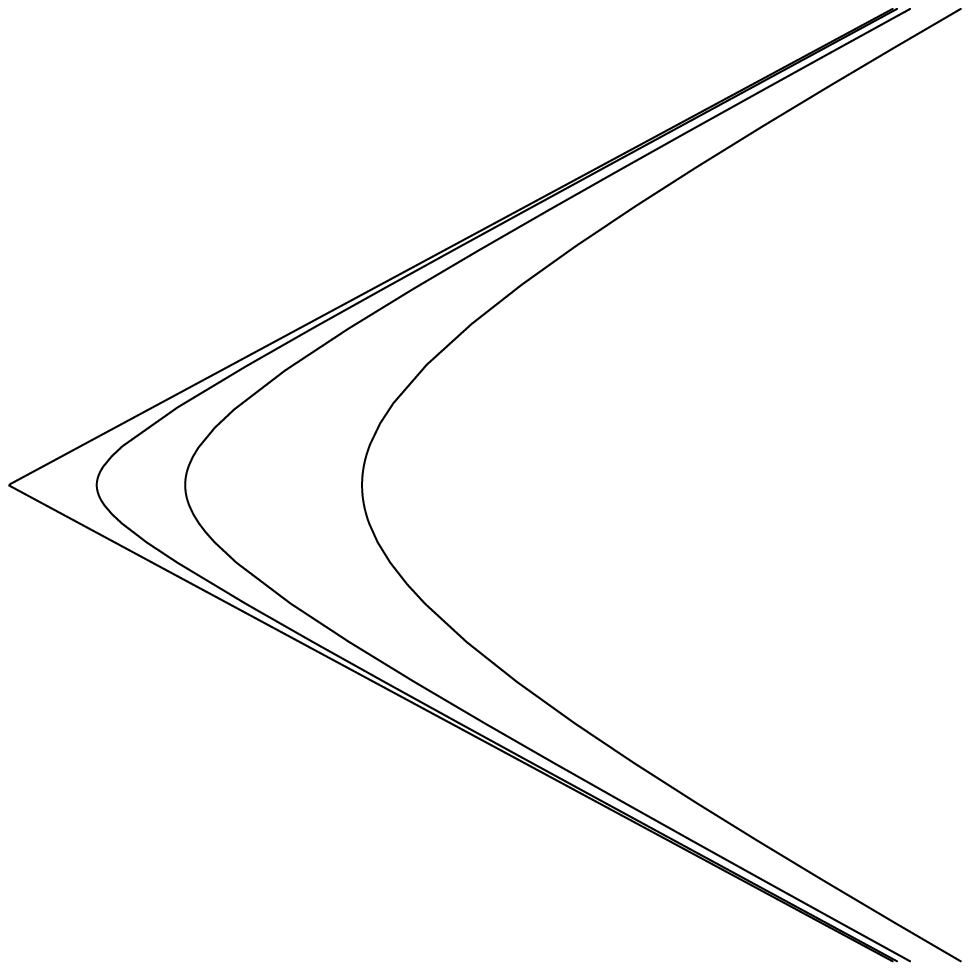,width=5cm}{A single
smoothed cone solution can be scaled down to provide a continuous
family of metrics which approach the cone. \label{ScaledCones}}

It turns out that a related dynamical system, given by the
Hamiltonian $H=(p_x^2+p_y^2)/2-x^2\, y^2/2$,
 \footnote{In order to cast this system in a form similar to ours (\ref{K2})
we transform first to the Lagrangian
$L=\(\dot{x}^2+\dot{y}^2+x^2\,y^2\)/2$ and then change variables
into $u=\log(x),\; v=\log(y)$ to obtain $L=\(e^{2 u}\, \dot{u}^2+
e^{2v}\, \dot{v}^2/v^2+\exp(2u+2v)\)/2$. In the $H=0$ sector we
may multiply $L$ (or $H$) by any lapse function, and we choose
$\exp(-u-v)$ to arrive at $L=\(e^{u-v}\, \dot{u}^2+ e^{v-u}\,
\dot{v}^2/v^2+\exp(u+v)\)/2$. This form still differs from
(\ref{K2}) but it is similar as it has a nearly canonical kinetic
term and an exponential potential.}
 was already analyzed in
quite a different physical setting as a model for critical
phenomena \cite{FLC:1998} (see also a higher dimensional
generalization \cite{Frolov:2006}). There one analyzes minimal
surfaces in a black hole background and one finds the (local)
critical solutions to be cones. The appearance of essentially the
same dynamical system (system of ODEs) in different physical
settings suggests that this dynamical system is in some sense the
minimal example for the physics of criticality including
self-similarity and critical exponents. Moreover, the same
Hamiltonian, apart from a sign change of the potential which is
inessential for current purposes, was already studied in
\cite{Savvidy:1982} in the context of Yang-Mills theories.

\section{Perturbations of the double cone}
\label{pert-section}


In this section we compute the spectrum of zero modes around the
double cone  (preserving the isometries of the black-hole
black-string system).

\presub {\bf Action}. Let us start by considering the following
ansatz, which is the most general one given the $U(1)_t \times
SO(D-2)_\Omega \times \IZ_{2,T}$ isometries of the black-hole
black-string system, where $\IZ_{2,T}$ stands for time reflection
$ t \to -t$
\begin{equation} ds^{2}=e^{2B_{\rho }}d\rho ^{2}+e^{2B_{\chi
}}(d\chi -fd\rho )^{2}+e^{2\Phi }dt^{2}+e^{2C}d\Omega _{D-3}^{2}
~. \label{def-fields}
\end{equation}%
All the fields are functions of $\rho$ and $\chi$ only. The ansatz
is ``most general'' in the sense that all of the Einstein
equations can be recovered by varying the gravitational action
with respect to these fields. From now on, in order to shorten the
notation we denote the derivative with respect to $\chi$ by a dot,
whereas the derivative with respect to $\rho$ is denoted by a
prime.

After some tedious computation the Lagrangian of the system can be
obtained
 \bea
 S &=& \int \exp \( \Psi + B_\rho + B_\chi \)\, d\rho\, d\chi ~ {\cal L} \non
 {\cal L} &=& K_1 + K_{2\rho} + K_{2\chi}- V ~,\eea
 where \bea
 -{1 \over D-3}\, K_1 &:=& \del C ~ \del\( \Psi + \Phi - C\)  \non
    &:=&  C' \, (\Psi'+\Phi'-C')  e^{-2 B_\rho} \non
    & & + \dot{C}\, (\dot{\Psi}+\dot{\Phi}-\dot{C}) \, \( e^{-2 B_\chi} + f^2\, e^{-2
    B_\rho}\)  \non
    & & + \( \dot{C}\, (\Psi'+\Phi'-C') + C'\, (\dot{\Psi}+\dot{\Phi}-\dot{C}) \)
     f \, e^{-2 B_\rho}   \non
 e^{2 B_\rho}\; K_{2\rho}
  &:=&  -2\, \dot{\Psi}\, (f\, \dot{f} + f' + f^2\, \dot{B}_\chi ) \non
  & & -2\, \Psi'\, (B'_\chi+ 2\,f\, \dot{B}_\chi ) \non
 & & +2 f (\dot{B}_\chi\, B'_\rho - B'_\chi\, \dot{B}_\rho) \non
 & & + 2\, \dot{f}\, (2\, B'_\chi-B'_\rho) -2\, f'\, (2\, \dot{B}_\chi - \dot{B}_\rho)\,  \non
 e^{2 B_\chi}\; K_{2\chi} &:= & -2\, \dot{\Psi}\, \dot{B}_\rho\,  \non
 V &:=& (D-3)(D-4)\, \exp(-2\, C) ~,
 \eea
 and we define \be
 \Psi := \Phi + (D-3)\, C  ~. \label{psidef}\ee
 The Lagrangian was divided into several parts as follows: $V$ is
the potential -- a part without derivatives and $K_1$ is a kinetic
term which is co-variant in the $(\rho,\chi)$ plane.  The rest of
the kinetic part was somewhat arbitrarily divided such that terms
with a $e^{-2 B_\rho}$ factor were collected into $K_{2\rho}$ and
the term with a $e^{-2 B_\chi}$ factor was denoted by $K_{2\chi}$.

\presub {\bf Gauge fixing}. This action is invariant under
reparameterizations of the $(\rho,\chi)$ plane (2 gauge
functions). We do not pretend to know an optimal gauge, but we
start by fixing \be
 f=0 \label{fix1} \ee
 as it seems to considerably simplify the equations. The
corresponding constraint comes from the equation of motion
 for $f\left( \rho ,\chi \right) $ and is given by \begin{eqnarray}
 0=\frac{1}{2}\left. {\delta S \over \delta f} \right|_{f=0}
&=& \bigl(-2\Psi^{\prime}\dot{B}_{\chi } + \dot{B}_{\chi }B_{\rho }^{\prime }-B_{\chi }^{\prime }\dot{B}_{\rho }  \notag \\
&&-(D-3)\bigl(\Phi ^{\prime }\dot{C}+\dot{\Psi}C^{\prime }-\dot{C}%
C^{\prime }\bigr)\bigr)e^{\Psi -B_{\rho }+B_{\chi }}  \notag \\
&&-\partial _{\rho }\left( \bigl(-\dot{\Psi}+\dot{B}_{\rho }-2
\dot{B}_{\chi }\bigl)e^{\Psi -B_{\rho }+B_{\chi }}\right)  \notag \\
&&-\partial _{\chi }\left( (2B_{\chi }^{\prime }-B_{\rho }^{\prime
})e^{\Psi -B_{\rho }+B_{\chi }}\right)  \notag \\
\label{constraint}
\end{eqnarray}%
Now substituting the gauge $f=0$ into the Lagrangian, we get%
\begin{gather}
L=\Bigl[-2 B_{\chi }^{\prime } \Psi ^{\prime } e^{-2B_{\rho }}
-(D-3) C^{\prime} \bigl( \Phi^{\prime }-C^{\prime}+\Psi^{\prime } \bigr)e^{-2B_{\rho }}  \notag \\
+(B_{\chi }\longleftrightarrow B_{\rho }
~,~``~^\prime~"\rightarrow
``~^{\cdot}~")-(D-3)(D-4)e^{-2C}\Bigr]e^{\Psi +B_{\rho }+B_{\chi
}} \label{Lagrangian}
\end{gather}

We choose to fix the remaining gauge freedom (beyond $f=0$) such
that the kinetic term with respect to $\chi$ (${\cal
O}(\del_\chi^{~2})~$ terms) in (\ref{Lagrangian}) is canonical (or
more precisely, field independent) \be
 B_\chi = \Psi + B_\rho + h \label{fix2} ~.\ee
where $h=h(\rho,\chi)$ will be fixed later. The associated
constraint is given by
\begin{eqnarray}
 0=\left. {\delta S \over \delta B_\chi} \right|_{{\rm fix}~B_\chi}
&=&\Bigl[2\dot{B}_{\rho }\dot{\Psi} +(D-3)\, \dot{C}\, (\dot{\Psi}+\dot{\Phi}-\dot{C}) \Bigr]e^{ -h} \notag \\
&&-\Bigl[ 2 (\Psi^{\prime }+B_{\rho }^{\prime }+h^{\prime
})\Psi^{\prime } + (D-3) C^{\prime } \( \Phi^{\prime }-C^{\prime
}+\Psi^{\prime }\) \Bigr]
e^{ 2\Psi + h}  \notag \\
&&-(D-3)(D-4)e^{ 2( \Psi- C +B_{\rho }) + h } + 2\partial _{\rho }\left( \Psi^{\prime } e^{ 2\Psi + h }\right)  
\label{bchi-constr}
\end{eqnarray}%

\presub {\bf The background}. The double-cone is a Ricci-flat
metric given by\footnote{See also (\ref{d-cone}).}
\begin{equation}
ds^{2}=d\rho ^{2}+\frac{\rho^2}{D-2}\left( d\chi ^{2}+\cos
^{2}\chi\; dt^{2}+\left( D-4\right) d\Omega
_{\mathbf{S}^{D-3}}^{2}\right) \label{eq_1}
\end{equation}%
As mentioned in the introduction, our objective here is to compute
the spectrum of perturbations around this metric, namely to solve
the Linearized Einstein equations around this background.

The gauge choice (\ref{fix1},\ref{fix2}) requires to redefine the
angle $\chi \rightarrow \widetilde{\chi }$. Choosing \be
h(\rho)=-(D-3)\ln{\( \rho\, \sqrt { \frac{D-4}{D-2} } \)} \ee
implies the following $\rho$-independent equation for
$\widetilde{\chi }$
 \begin{equation}
 \ln \frac{d\chi }{d\widetilde{\chi }}=\ln \cos \chi
\end{equation}%
The solution of this differential equation is given by%
\begin{equation}
\chi =\arctan (\sinh \widetilde{\chi })
\end{equation}%
where for simplicity we chose the constant of integration to be
zero. Note that while $\chi$ ranges over $[-\pi/2,\pi/2]$
$\widetilde{\chi}$ ranges over $[-\infty,+\infty]$. With this
redefinition at hand the solution (\ref{eq_1}) becomes
\begin{equation}
ds^{2}=d\rho ^{2}+\frac{\rho ^{2}}{D-2}\left( \frac{d\chi
^{2}+dt^{2}}{\cosh ^{2}\chi }+\left( D-4\right) d\Omega
_{\mathbf{S}^{D-3}}^{2}\right) ~,
\end{equation}%
here for simplicity of notation we omit tilde above $\chi$ and
denote $\widetilde{\chi}$ by $\chi$.

\presub {\bf Linearized equations}. Now we slightly perturb the
double-cone solution while keeping the gauge condition
(\ref{fix2}) unchanged, that is we set%
\begin{eqnarray}
B_{\rho } &=&B_{\rho }^{\left( 0\right) }+b_{\rho }= 0 + b_{\rho }
\non
B_{\chi } &=&B_{\chi }^{\left( 0\right) }+b_{\chi }=\ln \left( \frac{\rho }{%
\cosh \chi \sqrt{D-2}}\right) +b_{\chi } \non
 \Phi &=&\Phi^{\left( 0\right) }+\phi =\ln \left( \frac{\rho }{\cosh \chi
\sqrt{D-2}}\right) +\phi \non
 C &=&C^{\left( 0\right) }+c=\ln
\left( \rho \sqrt{\frac{D-4}{D-2}}\right) +c
\label{eq_5} \end{eqnarray}%
 together with the gauge condition derived from (\ref{fix2}) \be
 b_{\chi } =\psi +b_{\rho } \label{gauge_condition_2} ~,\ee
 where \be \psi:=(D-3)c+\phi \ee
(see also (\ref{psidef})).

Let us substitute equations (\ref{eq_5},\ref{gauge_condition_2})
into the $f$-constraint (\ref{constraint}) and $b_\chi$-constraint
(\ref{bchi-constr}). The zeroth order in small perturbations
vanishes in both cases, whereas the first order terms satisfy
the following equations%
\bea
 && (D-2)\dot{b}_{\rho }-\rho\, \dot{\psi}^{\prime} -\rho \tanh
\chi \( \psi^{\prime }-\phi^{\prime }+b_{\rho }^{\prime } \) =0
\label{constraint_f} \\ \non
 &&-2\rho ^{2} \psi^{\prime \prime }
  +2\rho \Bigl[\left( D-2\right) b_{\rho}^{\prime}-(D-1)\psi^{\prime} \Bigr]
  +\left( D-2\right) \Bigl[\dot{b}_{\rho }+\dot{\psi}-\phi\Bigr]\sinh
\left(2\chi \right) \notag \\
&& + 2\left( D-2\right) \left( D-3\right) \left( b_{\rho
}-c\right) =0 \label{constraint b_chi} \eea

Currently we are in the position to derive the equations of motion
for the rest of the fields. As a first step we substitute
(\ref{eq_5},\ref{gauge_condition_2}) into the Lagrangian
(\ref{Lagrangian}), expand
it and keep the quadratic part in the perturbations%

\begin{eqnarray}
\frac{\cosh ^{2}\chi }{\rho ^{D-3}}L^{\left( 2\right) }&=&
-\frac{\rho ^{2}}{\left( D-2\right) }\Bigl[\left( 3 D-10\right)
(D-3)c'^{~2}+6\phi ^{\prime} \psi' -4\phi'^{~2}+2b_{\rho}\,'\, \psi'\Bigr] \notag \\
&&-\Bigl[(D-3)
\dot{c}(\dot{\psi}-\dot{c}+\dot{\phi})+2\dot{b}_{\rho
}\dot{\psi}\Bigr]\cosh ^{2}\chi
-4\rho \(2\psi^{\prime}+b_{\rho}^{\prime}\) \psi   \notag \\
&&-2(D-1)\psi ^{2} -2(D-3)\( \psi - c +b_{\rho }\) ^{2}
\label{L_2}
\end{eqnarray}

As a result, the equations of motion for the fields $b_{\rho }$,
$c$ and $\phi$ are given respectively by

\begin{eqnarray}
 && 2\left[ (D-3)b_{\rho }-2\psi + \phi \right]=\frac{\left(
3D-5\right) }{ \left( D-2\right) }\rho\, \psi^{\prime} +\frac{\rho
^{2}}{\left( D-2\right) }\psi^{\prime \prime} +\cosh ^{2}\chi~
\ddot{\psi} \non \non
 &&\rho \left[ \left(
D-1\right)\(3\psi^{\prime}-c^{\prime}\) -(D-3)b_{\rho
}^{\prime}\right]
+\rho ^{2}\( 3\psi^{\prime\prime}-c^{\prime\prime}+b_{\rho}^{\prime\prime} \)  \notag \\
&&+\left( D-2\right) \cosh ^{2}\chi \( \ddot{\psi}- \ddot{c}+\ddot{b_{\rho}}\)  \notag \\
&=&2\left( D-2\right) \Bigl( c-2\psi+\phi+(D-4)b_{\rho} \Bigr)
\non \non
0&=&\rho \left[ (D-1)(3\psi^{\prime}-\phi^{\prime})-(D-3) b_{\rho
}^{\prime}\right]
+\rho ^{2}\( 3\psi^{\prime\prime}-\phi^{\prime\prime}+b_{\rho}^{\prime\prime}\) \notag \\
&&+\left( D-2\right) \cosh ^{2}\chi \( \ddot{\psi}-
\ddot{\phi}+\ddot{b_{\rho}} \) -2\left( D-2\right) (D-3)\( b_{\rho
}-c\) \label{eom}
\end{eqnarray}

Let us denote the constraint (\ref{constraint_f}) coming from the
gauge fixing condition $f=0$ by $A$, and by $B$ the constraint
(\ref{constraint b_chi}) coming from the gauge condition
(\ref{gauge_condition_2}). One finds that the following relations
hold on the solutions of
equations of motion (\ref{eom})%
\begin{eqnarray}
\partial _{\chi }A\left( \rho ,\chi \right) &=&-\frac{\rho \partial _{\rho
}B\left( \rho ,\chi \right) }{2\left( D-2\right) \cosh ^{2}\chi }  \notag \\
\partial _{\chi }B\left( \rho ,\chi \right) &=&2\left( D-2\right) A\left(
\rho ,\chi \right) +2\rho \partial _{\rho }A\left( \rho ,\chi
\right) +2\tanh \chi ~ B\left( \rho ,\chi \right)
\end{eqnarray}%
It means, that once the constraints are satisfied for some value
of the coordinate $\chi $ they will vanish identically for any
$\chi$.

\subsection{Solving the equations}

We proceed to solve first the equations of motion (\ref{eom}), and
then later select those solutions which satisfy also the
constraints (\ref{constraint_f},\ref{constraint b_chi}).

In order to solve the equations of motion (\ref{eom}) we
attempt to separate the variables through the following general ansatz%
\begin{equation}
\overrightarrow{X}\left( \rho ,\chi \right) =%
\begin{pmatrix}
b_{\rho }^{\left( 1\right) } \\
\phi ^{\left( 1\right) } \\
c^{\left( 1\right) }%
\end{pmatrix}%
=\overrightarrow{X}^{l}\rho ^{s}P_{l}\left( \tanh \chi \right)
\label{Xansatz}
\end{equation}%
where $l$ is any non-negative integer, $\overrightarrow{X}^{l}$
are unknown constant vectors and $P_{l}\left( \tanh \chi \right)$
are the Legendre polynomials which can be shown to satisfy the
following differential equation%
\begin{equation}
\cosh ^{2}\chi\, \frac{d^{2} }{d\chi ^{2}} P_{l}\left( \tanh \chi \right)%
+l\left( l+1\right) P_{l}\left( \tanh \chi \right) =0
\label{Legendre2}
\end{equation}%
The description ``general ansatz'' needs some explanation. Once we
add a sum over $l$ to the r.h.s of (\ref{Xansatz}) this becomes
the most general decomposition since $\rho^s$ and $P_l$ form a
complete basis of functions. The question is whether the variables
separate in (\ref{eom}), thus allowing to omit the sum. {\it A
priori} the $SO(3)$ isometry of the background tells us that the
angular coordinates can be separated into spherical harmonic
functions. In general such functions are labelled by $l,m$ but the
$U(1)_t$ isometry implies that only $m=0$ terms contribute. If our
fields included only scalars then the equations would be
guaranteed to separate into the scalar spherical harmonics
$Y_{l0}$, namely the Legendre polynomials. In our case the
perturbation includes also tensor modes (with respect to the
$\IS^2$), but still direct inspection
 \footnote{$\chi$ appears in
(\ref{eom}) only through the combination
$\cosh^2{\chi}~\del^2_\chi$ and after use of (\ref{Legendre2}) all
$\chi$ dependence disappears.}
 confirms that equations (\ref{eom}) separate
under the ansatz (\ref{Xansatz}).

As a result we get the following set of algebraic equations%
\begin{equation}
\left[ s\left( s-1\right) E_{2}+sE_{1}-l\left( l+1\right) K\right]
\overrightarrow{X}=V\overrightarrow{X}
\end{equation}%
where we have defined the following $3 \times 3$ constant matrices%
\begin{eqnarray}
E_{1} &=&%
\begin{bmatrix}
\room -\left( D-3\right)  & \room2\left( D-1\right)  & \room3(D-3)\left( D-1\right)  \\
\room 0 & \room\left( 3D-5\right)  & \room\left( 3D-5\right) (D-3) \\
\room -(D-3) & \room3\left( D-1\right)  & \room\left( D-1\right)
\left( 3D-10\right)
\end{bmatrix}%
;  \notag \\
E_{2} &=&%
\begin{bmatrix}
\room1 & \room2 & \room3(D-3) \\
\room0 & \room1 & \room(D-3) \\
\room1 & \room3 & \room\left( 3D-10\right)
\end{bmatrix}
;  \notag \\
K &=&\left( D-2\right)
\begin{bmatrix}
\room1 & \room0 & \room(D-3) \\
\room0 & \room1 & \room(D-3) \\
\room1 & \room1 & \room\left( D-4\right)
\end{bmatrix}%
;  \notag \\
V &=&2\left( D-2\right)
\begin{bmatrix}
\room(D-3) & \room0 & \room-(D-3) \\
\room(D-3) & \room-1 & \room-2(D-3) \\
\room(D-4) & \room-1 & \room\left( 7-2D\right)
\end{bmatrix}%
;
\end{eqnarray}%
The spectrum of $s$ is determined from the following
characteristic equation%
\begin{equation}
Det\left[ s\left( s-1\right) E_{2}+sE_{1}-l\left( l+1\right)
K-V\right] =0
\end{equation}%
After some tedious algebraic rearrangement, one
can simplify the above equation and get%
\begin{eqnarray*}
s^{2}+\left( D-2\right) s-\left( l+2\right) \left( l+1\right)
\left(
D-2\right)  &=&0 \\
s^{2}+\left( D-2\right) s-\left( l+2\right) \left( l-1\right)
\left(
D-2\right)  &=&0 \\
s^{2}+\left( D-2\right) s-l\left( l-1\right) \left( D-2\right)
&=&0
\end{eqnarray*}%
The solutions are given by\footnote{%
Indices $1$,$3$ and $5$ correspond to upper sign, whereas indices
$2$,$4$
and $6$ to lower one.}%
\begin{eqnarray}
s_{1,2} &=&\frac{1}{2}\left( 2-D\pm \sqrt{\left( D-2\right) \left(
4l^{2}+12l+D+6\right) }\right)   \notag \\
s_{3,4} &=&\frac{1}{2}\left( 2-D\pm \sqrt{\left( D-2\right) \left(
4l^{2}+4l+D-10\right) }\right)   \notag \\
s_{5,6} &=&\frac{1}{2}\left( 2-D\pm \sqrt{\left( D-2\right) \left(
4l^{2}-4l+D-2\right) }\right)   \label{s}
\end{eqnarray}

whereas the corresponding eigenvectors are%
\begin{eqnarray}
\overrightarrow{X}_{1,2}^{l} &=&%
\begin{bmatrix}
\room(l+2)\left[ 4-D\pm \sqrt{\left( D-2\right) \left(
4l^{2}+12l+D+6\right) }\right] \\
\room2l-D+6\mp \sqrt{\left( D-2\right) \left( 4l^{2}+12l+D+6\right) } \\
\room2(l+2)%
\end{bmatrix}
\notag \\ \notag \\ \notag \\
\overrightarrow{X}_{3,4}^{l} &=&%
\begin{bmatrix}
\room 4(D-2)(D-3)(l-1)(l+2) \\
\room \left( D-3\right) \left[ D-2\pm \sqrt{\left( D-2\right)\left( 4l^{2}+4l+D-10\right) }\right]^2  \\
\room -4(D-2)(l-1)(l+2)
\end{bmatrix}
\notag \\ \notag \\ \notag \\
\overrightarrow{X}_{5,6}^{l} &=&%
\begin{bmatrix}
\room (l-1)\left[ 4-D\pm \sqrt{\left( D-2\right) \left(
4l^{2}-4l+D-2\right)
}\right] \\
\room 2l+D-4\pm \sqrt{\left( D-2\right) \left( 4l^{2}-4l+D-2\right) } \\
\room 2(l-1)
\end{bmatrix}
\end{eqnarray}

It turns out that $\overrightarrow{X}_{3,4}$ satisfy the
constraint equations (\ref{constraint b_chi},\ref{constraint_f})
as well, and thus they are part of the perturbation spectrum. On
the other hand, $\overrightarrow{X}_{1,2}$ and
$\overrightarrow{X}_{5,6}$ do not satisfy the constraints
independently. Since mixing of different modes is allowed for
those values of $s$ which are degenerate, one concludes that in
order to find out other possible solutions of the perturbation
spectrum we need to take a superposition of
$\overrightarrow{X}_{1,2}$ and $\overrightarrow{X}_{5,6}$
corresponding to the same values of $s$ and then check whether
such a combination can satisfy the constraints.

According to (\ref{s}) one can see, that in order to make the
powers of $\rho $ equal we should consider the
following superpositions%
\begin{eqnarray}
\overrightarrow{X}\left( \rho ,\chi \right) &=&\rho ^{s_{1}}\left[
\overrightarrow{X}_{5}^{l+2}P_{l+2}\left( \tanh \chi \right) +F%
\overrightarrow{X}_{1}^{l}P_{l}\left( \tanh \chi \right) \right]  \notag \\
\overrightarrow{X}\left( \rho ,\chi \right) &=&\rho ^{s_{2}}\left[
\overrightarrow{X}_{6}^{l+2}P_{l+2}\left( \tanh \chi \right) +G%
\overrightarrow{X}_{2}^{l}P_{l}\left( \tanh \chi \right) \right]
\end{eqnarray}%
where $F$and $G$ are constants to be determined. Substituting
these linear superpositions into constraints equations, we find
that the constraint can be satisfied by taking $F=G=-1$, and we
have another family of solutions.

In summary, the full perturbation spectrum
is given by%
\begin{equation}
\fbox{$%
\begin{array}{c}
s_{\pm}^t=\frac{1}{2}\left( 2-D\pm \sqrt{\left( D-2\right) \left(
4l^{2}+12l+D+6\right) }\right)  \\
s_{\pm}^s=\frac{1}{2}\left( 2-D\pm \sqrt{\left( D-2\right) \left(
4l^{2}+4l+D-10\right) }\right)
\end{array}%
$} \label{spectrum}
\end{equation}
where\footnote{``t" stands for ``tensor" and ``s" for ``scalar".}
$l\geq 0$.

For $s_{\pm}^t$ the modes are given by \begin{eqnarray}
\overrightarrow{X}_{+}^t\left( \rho ,\chi \right) &=&\rho
^{s_{+}^t}\left[
\overrightarrow{X}_{5}^{l+2}P_{l+2}\left( \tanh \chi \right) -%
\overrightarrow{X}_{1}^{l}P_{l}\left( \tanh \chi \right) \right]   \notag \\
\overrightarrow{X}_{-}^t\left( \rho ,\chi \right)  &=&\rho
^{s_{-}^t}\left[
\overrightarrow{X}_{6}^{l+2}P_{l+2}\left( \tanh \chi \right) -%
\overrightarrow{X}_{2}^{l}P_{l}\left( \tanh \chi \right) \right]
\eea
 while for $s_{\pm}^s$ they are given by \be
\overrightarrow{X}_{\pm}^s\left( \rho ,\chi \right)  =\overrightarrow{X}%
_{3,4}^{l} ~\rho ^{s_{\pm}^s}P_{l}\left( \tanh \chi \right) \ee

According to the ``$s_+$ prescription'' boundary condition (below
(\ref{irrelevant})), we should eliminate all the $s_-$ modes. For
$s^s(l=0)$ this prescription is ambiguous, but after studying the
$l=0$ sector in detail in the next section we will conclude that
still  the b.c. reduce the dimension of the solution space from 2
to 1.

In addition, another mode should be eliminated from the above
perturbation spectrum, namely $\overrightarrow{X}_{+}^s\left( \rho
,\chi \right)$ for $s_{+}^s(l=1)=0$, since it corresponds to a
residual gauge of an infinitesimal shift in $\chi$ coordinate.

Altogether it can be seen that except for $s^s_+(l=0)$ all other
$s_+$ (physical) modes are positive, thus satisfying our stability
criterion (\ref{stability}).

\section{Non-linear spherical perturbations}
\label{non-pert-section}


In this section we obtain the qualitative features of the dynamics
of the full non-linear perturbations in the ``spherical'' sector
($l=0$, namely preserving all isometries).

\presub {\bf Action}. The most general D-dimensional metric with
$SO(m+1) \times SO(n+1)$ isometry, $D=m+n+1$ is \be
 ds^2=e^{2\, B_\rho}\, d\rho^2 + e^{2\, A}\, d\Omega^2_m + e^{2\, C}\,
 d\Omega^2_n  ~,\label{ansatz-brho} \ee
 where all three functions $B_\rho,\, A$ and $C$ depend on $\rho$
 only, $d\Omega^2_m$ and $d\Omega^2_n$ are the standard metrics on
the $m$ and $n$ spheres, and there is a reparameterization gauge
freedom $\rho \to \rho'(\rho)$. For applications to the black-hole
black-string system we need only the case $m=2$, which represents
the $\chi,t$ 2-sphere while the $n$-sphere is the angular sphere.

The action is \bea
 S &=& \int d\rho \, e^{m\,A + n \, C}\, e^{B_\rho}
 \[ e^{-2\, B_\rho}\, K- \wV \] \non
  K &=& {m\, n \over (D-1)}\, (A'-C')^2- \(1-{1 \over D-1}\)\, (m\, A' +
 n \, C')^2 = \non
  &=& -m(m-1)\, A'^2 - n(n-1)\, C'^2 - 2mn\, A'\, C' \non
 \wV &=& m\, (m-1)\, e^{-2\, A} +n\, (n-1)\, e^{-2\, C}
\eea
 where a prime denotes a derivative with respect to $\rho$
(the overall sign was chosen such that $S=-\int \sqrt{g}\, R$).

The system enjoys a scaling symmetry
 \footnote{
The variation of the action due to an infinitesimal symmetry
operation is $\delta_\al S= S$. Usually one considers symmetries
which vary the action only by a boundary term thus defining a
conserved current, which is absent in this case. Still this
variation is enough to guarantee that the equations of motion are
satisfied. Indeed, assuming we have a solution for the equations
of motion $\delta S/\delta \phi = 0$ (where $\phi$ is a collective
notation for the fields) then their variation vanishes as well
$\delta_\al\, \delta S/\delta \phi = \delta (\del_\al S)/\delta
\phi = \delta S/\delta \phi = 0$.}
 \be
 ds^2 \to e^{2 \al} ds^2 \ee
 namely \bea
 B_\rho &\to& B_\rho + \al \non
  A &\to& A + \al \non
  C &\to& C + \al ~. \label{scaling}\eea

It is convenient to fix the gauge such that the kinetic term is
canonical (more precisely, its prefactor is field independent),
namely \be
 B_\rho= m\, A + n\, C \label{gauge-cond} ~.\ee
 The ansatz reads
 \be
 ds^2=e^{2m\, A + 2n\, C}\, d\rho^2 + e^{2\, A}\, d\Omega^2_m + e^{2\, C}\,
 d\Omega^2_n  ~,\label{ac-ansatz} \ee
while the action becomes \bea
  S &=& \int d\rho\,
   \[ K- V \] \\
   V &:=& \exp(2m\, A+ 2n\, C)\, \wV
   \equiv m\, (m-1)\, e^{2(m-1)\, A+2 n\, C} +n\, (n-1)\, e^{2m\, A+ 2(n-1)\,C}~, \nonumber
   \eea
   and it is supplemented by the constraint \be
 0=H:=K+V \label{Hconstraint} ~.\ee

\presub {\bf Change of variables}. It is convenient to make the
following field re-definitions. First one makes a linear
re-definition that simplifies the potential term \bea
 u &:=& 2(m-1)\, A + 2n\, C + \log\(m(m-1)\) \non
 v &:=& 2 m\, A + 2(n-1)\, C + \log\(n(n-1)\) ~. \label{uv} \eea
 The potential and kinetic terms become \bea
 V &=& e^u + e^v \non
 K &=&  {1 \over 4(D-1)}\, \[m\, n \, (u'-v')^2- {1 \over (D-2)}\, (m\, u' +  n \, v')^2 \]=
 \non
   &=&  {1 \over 4(D-2)} \[ m(n-1)\, u'^2 + n(m-1)\, v'^2 - 2mn\, u'\, v'
 \] \label{K2} \eea

This is a system with two degrees of freedom and a potential which
is a sum of two exponentials. At first it appears to be similar to
a Toda system, but actually it is probably not integrable for the
following reason: the spring potential in a Toda system is of the
form $e^x-x$ (where $x$ measures the deviation from equilibrium
length), and while the linear term often cancels due to the mass
being acted on by two springs, one from each side, here there are
only two springs for two masses, and thus our system which lacks a
linear term is not of a Toda form.

Still it is convenient to make the following ``Toda inspired''
change of variables \be \begin{array}{cc}
 X_1 := e^u  & \hspace{1cm} P_1 := u' \\
 X_2 := e^v &  \hspace{1cm} P_2 := v' \label{XP} \end{array} ~, \ee
 where the notation should not be mistaken to imply that the $X$'s
and $P$'s are conjugate variables. The equations of motion read
\be
\begin{array}{cc}
 X_1' = P_1\, X_1  & \hspace{1cm}
  P_1' = 2 \( (1-{1 \over m})\, X_1 + X_2 \)  \\
 X_2' = P_2 X_2     & \hspace{1cm}
  P_2':= 2 \(  X_1 + (1-{1 \over n})\, X_2 \) \label{EOM-XP} \end{array} ~, \ee
and the constraint (\ref{Hconstraint}) becomes \be
 0 = {1 \over 4(D-1)}\, \[m\, n \, (P_1-P_2)^2- {1 \over (D-2)}\, (m\, P_1 +  n \, P_2)^2 \]
  +X_1+X_2 \label{constraint2} ~.\ee
The $(X_i,P_i), ~i=1,2$ variables also provide a convenient
realization of the scaling symmetry (\ref{scaling}) \bea
 X_i &\to& e^{2\, \hat{\al}}\, X_i \non
 P_i &\to& e^{\hat{\al}}\, P_i \non
 \rho  &\to& e^{-\hat{\al}}\, \rho \label{scaling-PX} \eea
where $\hat{\al}$ is related to $\al$ in (\ref{scaling}) through
$\hat{\al} := (D-2)\, \al$.

The double cone metric in the $X,P$ variables is found either by
transforming (\ref{d-cone}) according to the changes of variables
(\ref{uv},\ref{XP}) or by solving directly the equations of motion
(\ref{EOM-XP}) subject to scaling (\ref{scaling-PX}) invariance. It
is given by \bea
 P_1 &=& P_2  = -{2 \over \rho} \non
 X_1 &=& { m \over D-2}\,  {1 \over \rho^2} \non
 X_2 &=& { n \over D-2}\,  {1 \over \rho^2} \label{dcone-PX} \eea

Next we transform to \bea
 X^+ &:=& X_1 + X_2 \non
 X^- &:=& {1 \over m}\, X_1 - {1 \over n}\, X_2 \non
 P^+ &:=& m\, P_1 + n\, P_2 \non
 P^- &:=& P_1 - P_2 \label{XP+-} ~.\eea
 To arrive at these definitions we first considered the 5d case
$m=n=2$ where by symmetry it is useful to transform to $X_1 \pm
X_2,\, P_1 \pm P_2$ and then we generalized to arbitrary $m,n$
paying attention to the form of the kinetic energy (\ref{K2}).
 The equations of motion read \bea
X^+\,' &=& {1 \over m+n} \( X^+\, P^+ + mn\, X^-\, P^- \) \non
 X^-\,' &=& {1 \over m+n} \( X^+\, P^- + X^-\, P^+ + (n-m)\, X^-\, P^- \) \non
 P^+\,' &=& 2 (D-2)\, X^+ \non
 P^-\,' &=& -2\, X^- \eea

\presub {\bf Fixing the scaling symmetry}. Now comes the crucial
step in the analysis, which will allow the qualitative solution of
the dynamical system. The phase space consists of 4 variables
$X_i,\,P_i$ constrained by (\ref{constraint2}) and hence it is 3d.
However, dynamical systems in 3d can be quite involved, and we would
not know how to analyze this system. Fortunately the scaling
symmetry (\ref{scaling},\ref{scaling-PX}) can be used to reduce the
problem to a 2d phase space, where the number of qualitative
possibilities is quite limited and a full qualitative analysis is
possible.

The idea is to fix the symmetry by choosing a 2d cross section of
the phase space which is transverse to the symmetry orbits. Then
we supplement the infinitesimal $\rho$-evolution by an
infinitesimal symmetry operation such that we always remain on the
2d cross-section, thereby reducing the problem to a 2d phase
space.

In practice we fix the symmetry as follows. Being transverse to a
scaling symmetry means introducing an arbitrary scale. We choose the
following condition \be
 X^+ =1 \label{fix} ~.\ee

In order to define the reduced ``scaling compensated'' evolution we
introduce a condensed notation for the phase space variables \be
 Y^i=(X_1,X_2,P_1,P_2) ~.\ee
 The scaling transformation is given by $Y^i \to e^{q^i \hal}\,
 Y^i$ where \be
 q^i=(2,2,1,1) ~.\ee
We denote the functions on the r.h.s of (\ref{EOM-XP}) by $f^i(Y^j)$
such that \be
 Y^i\,'=f^i(Y^j) ~. \label{EOM-Y} \ee
Note that since $\rho$ carries scaling charge $q_\rho=-1$ $f^i$
carry charge $q^i+1$.

We need to distinguish the reduced evolution parameter $\rho_R$ from
$\rho$ since scaling acts on $\rho$ as well (\ref{scaling-PX}). Now
we can define the reduced phase space trajectory
$Y^i_R=Y^i_R(\rho_R)$ by \be
 \cdr Y^i_R = f^i - q^i\, {f^+ \over 2\, X^+}\, Y^i \label{EOM-YR} \ee
where $f^+/(2\, X^+)$ is the compensating infinitesimal scaling
parameter and is defined such $\cdr
 X^+_R=0$. For clarity, we write down the definition of $f^+$
 explicitly \be
f^+(Y^i_R):={1 \over m+n}\, (P^+_R + m n\, X^-_R\, P^-_R)~. \ee

We parameterize the reduced 2d phase space by $(X,\theta)$ where \be
 X \equiv X^-_R ~,\ee
and $\theta$ is a hyperbolic angle which parameterizes the hyperbola
in $P$ space given by the constraint (\ref{constraint2}) together
with the condition (\ref{fix}), namely \bea
 \cosh{\theta} &=& -{1 \over 2\sqrt{(D-2)\,(D-1)}} ~ P^+_R \non
 \sinh{\theta} &=& -\half \sqrt{{m n \over D-1}} ~ P^-_R \label{theta} ~.\eea
The signs above depend on conventions: the sign in the definition of
$\cosh(\theta)$ is related to setting the direction of the flow
towards the tip, and the sign in the definition of $\sinh(\theta)$
sets the sign of $\theta$.
 \emph{Our (final) expression for the reduced equations is} \bea
 \cdr X &=& -{2 \over \sqrt{m n}}\, \sinh{\theta}\, (1+ n\, X )\,
 (1-m\, X) \non
 \cdr \, \theta &=&  \sqrt{D-2}\, \sinh{\theta} + \sqrt{m n} ~ \cosh{\theta}~ X ~,
 \label{2deq} \eea
 where we chose to rescale $\rho_R \to \rho_R/\sqrt{m+n}$.

Given a solution of (\ref{2deq}), or equivalently functions
$Y^i_R(\rho_R)$ satisfying (\ref{EOM-YR}), we still need to
\emph{uplift} it to a solution of (\ref{EOM-Y}). First we integrate
for the scale factor evolution $\hal(\rho_R)$ \be
 \cdr \hal = {f^+(y^i_R(\rho_R)) \over 2 X^+_R} = \half\, f^+(y^i_R(\rho_R)) \label{hal-lift} ~.\ee
 Then we define the uplifted trajectory $Y^i(\rho)$ by \bea
 Y^i = e^{q^i \hal}\, Y^i_R \non
 d\rho=e^{-\hal}\, d\rho_R ~. \label{uplift} \eea
A direct computation confirms that with this definition
(\ref{EOM-Y}) is satisfied  $${d \over d\rho}\, Y^i= e^{\hal}\,
\cdr e^{q^i \hal}\, Y^i_R =e^{(q^i+1) \hal}\, (\cdr Y^i + q^i \cdr
\hal ) = e^{(q^i+1)}\, f^i(Y^j_R)= f^i(Y^j)~,$$
 where in the first equality we used (\ref{uplift}), in the second to last we used
(\ref{EOM-YR},\ref{hal-lift}) and finally in the last equality we
used the fact that $f^i$ has charge $q^i+1$ with respect to scaling.

\subsection{Analysis of the reduced 2d phase space}

The analysis of the qualitative form of the reduced 2d phase space
(\ref{2deq}) proceeds by determining the domain, equilibrium
points, their nature (at linear order) and a separate analysis of
the behavior at infinity.

\presub {\bf Domain}. From the definition of $X_1,\, X_2$
(\ref{XP}) it is evident that they are both positive. Given the
gauge fixing condition (\ref{fix}) and the change of variables
(\ref{XP+-}) we find that the domain is restricted to the strip
\be
 -{1 \over n} \le X \le {1 \over m} \label{domain} ~.\ee
The boundary is strictly speaking not part of the domain, but since
the equations (\ref{2deq}) continue smoothly to the boundary while
$\cdr X$ vanishes there, we can join it to the domain and allow for
equalities in (\ref{domain}).

\presub {\bf Equilibrium points}. There are 3 finite equilibrium
points \be \begin{array}{cc}
 X_0=0 & \theta_0=0 \\
 X_1={1 \over m} ~~& ~~\tanh(\theta_1) = -\sqrt{{m n \over D-2}}\, {1 \over m}
 \\
 X_2=-{1 \over n} ~~& ~~\tanh(\theta_2) = \sqrt{{m n \over D-2}}\, {1 \over n}
 \end{array} ~,
 \label{equilib} \ee
two of which are on the boundary. The point $(0,0)$ corresponds to
the double cone (\ref{dcone-PX}), which is indeed a fixed point of
the scaling symmetry. The role of the other points will become clear
below.

\begin{figure}[t!]
\centering \noindent
\includegraphics[width=7cm]{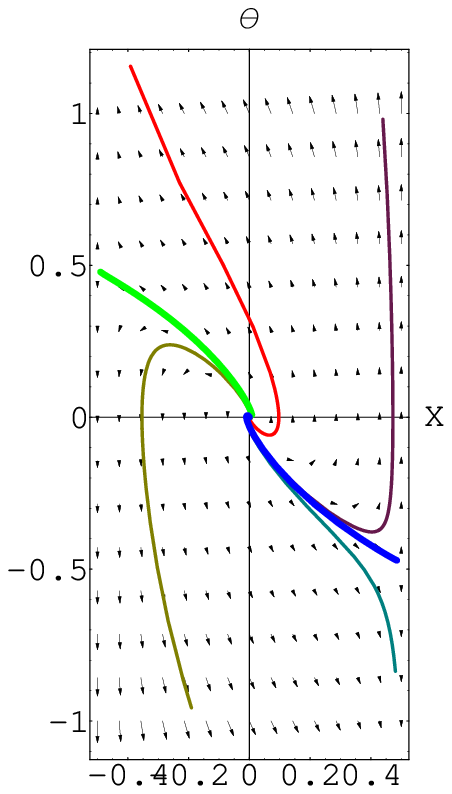}
\caption[]{The phase space (for $m=n=2$). There is one interior
equilibrium point at $(0,0)$ which represents the double cone. For
$D<10$ it is focal repulsive but the spirals cannot be seen in the
figure since their log-period is too large, while for $D\ge 10$ it
is nodal repulsive, and hence the figure represents the whole range
of dimensions. There are two finite equilibria on the boundary at
$(1/m,\theta_1)$ and $(-1/n,\theta_2)$ which are saddle points. Each
one has a critical curve which approaches it (heavy line). These two
special trajectories denote the smoothed cones, where either $\IS^n$
or $\IS^m$ shrinks smoothly while the other one stays finite.
Finally there are two attractive equilibria at infinity at
$(1/m,-\infty)$ and $(-1/n,+\infty)$, and the thin lines
trajectories represent generic trajectories which end at these
infinite equilibrium points.} \label{phase-space5d}
\end{figure}

\presub {\bf Linearized analysis of equilibrium points}. The
linearized dynamics around the an equilibrium point $(X_q,\theta_q)$
is given by the $2 \times 2$ matrix $L$ \be
 \cdr \[ \begin{array}{c} \delta X \\ \delta \theta \end{array} \]
 = L
\[ \begin{array}{c} \delta X \\ \delta \theta \end{array} \] ~,\ee
 where $\delta X:=X-X_q,~ \delta \theta:=\theta-\theta_q$.
We remind the reader of the classification of equilibrium points
\begin{itemize}
 \item If the eigenvalues of $L$ are complex (and necessarily self-conjugate
 since $L$ is real) then
trajectories circle $(X_q,\theta_q)$ or spiral around it, and the
point is called a \emph{focal point}.
 \item If the eigenvalues of $L$ are real and of opposite sign then
$(X_q,\theta_q)$ is called a \emph{saddle point} and it is unstable
in the sense that only if the initial conditions are fine-tuned to
lie on a specific trajectory will the evolution flow into this
point.
 \item If the eigenvalues of $L$ are real and of the same sign then
$(X_q,\theta_q)$ is called a \emph{nodal point}. If they are both
positive then the point is called \emph{repulsive}, while for
negative it is called \emph{attractive}. Naturally, the adjectives
repulsive and attractive interchange under inversion of the flow
(time reversal). A focal point may be called repulsive (or
attractive) as well if the real parts of $L$'s eigenvalues are all
positive (or negative).
\end{itemize}

For the double-cone $(X,\theta)=(0,0)$ we find \be
 L_0=\[ \begin{array}{cc} 0 & ~~~~-2/ \sqrt{m n} \\
   \sqrt{m n} & ~~~~ \sqrt{D-2} \end{array} \] ~.\label{Ldcone} \ee
The eigenvalues of this matrix are \be
 \lambda_\pm = {1 \over 2}\, \( \sqrt{D-2} \pm \sqrt{D-10} \) \ee
 Let us make several observations
 \bi
 \item The eigenvalues depend only on $D$ and not on $m,n$
 separately.
 \item $D=10$ is a critical dimension, as found in
 \cite{TopChange}. \bi
 \item \emph{For $D<10$} the eigenvalues are complex and we have a  \emph{repulsive focal
 point}.
 \item \emph{For $D \ge 10$} the eigenvalues are real and we have a \emph{repulsive nodal
 point}.
 \item For the critical value $D=10$ the eigenvalues degenerate,
but $L$ is not proportional to the identity matrix, but rather $L$
can be brought by a similarity transformation to the form  \be
 \[ \begin{array}{cc} \sqrt{2} & ~~1 \\
  0 &  ~~ \sqrt{2} \end{array} \] ~.\label{Ldcone10D} \ee \ei
 \item In the $D \to \infty$ limit we have $\lambda_+ \simeq
\sqrt{D},~ \lambda_- \simeq 2/\sqrt{D}$.
 \item  The eigenvectors which correspond to the eigenvalues
 $\lambda_\pm$ are $[2, ~-\lambda_\pm\, \sqrt{m n} ]$.
 \ei

At the equilibrium point $(X_1,\theta_1)$ we find \be
 L_1= \cosh{\theta_1}\, \[ \begin{array}{cc} {-2(D-1) \over m \sqrt{D-2}} & ~~~~0 \\
  \sqrt{m n} & ~~~~\sqrt{D-2} - {n \over m\sqrt{D-2}} \end{array} \] ~.\label{Lsmooth-cone} \ee
Similarly $L_2$ is obtained by substituting $\theta_1
\leftrightarrow \theta_2,~ m \leftrightarrow n$. Since $\det{L}<0$
we see that \emph{at both $(1/m,\theta_1)$ and $(-1/n,\theta_2)$
we find a saddle point}. The positive direction (eigenvector of
the positive eigenvalue, defining the repulsive direction) is $[0,
~1 ]$ in both cases, namely along the boundary.

\presub {\bf Behavior at infinity}. There are two attractive
equilibrium points at infinity \be \begin{array}{cc}
 X_3={1 \over m} ~~ & ~~\theta_3 = -\infty \\
 X_4=-{1 \over n} ~~& ~~\theta_4 = +\infty \end{array} ~.
 \label{equilib-inf} \ee
 $(X_3,\theta_3)$ is attractive since for $\theta<\theta_1$ we
 have $dX/d\rho_R>0, ~d\theta/d\rho_R<0$. There are no other
fixed points at $\theta=-\infty$ but with $-1/n < X < 1/m$ since
$\lim_{\theta \to -\infty} dX/d\theta < 0$. Moreover, close to
$(X_3,\theta_3)$ the trajectories are found to asymptote
exponentially fast to $X_3$, namely $|\delta X| \simeq \exp(-const\,
\theta)$. Analogous properties hold for the attractive equilibrium
point $(X_4,\theta_4)$.

\presub {\bf The whole picture}. Having found all the ingredients
above we may assemble them into a complete phase diagram, figure
\ref{phase-space5d}, which is discussed in its caption.

The trajectories which end at $(X_1,\theta_1)$ and
$(X_2,\theta_2)$ represent the smoothed cones. This is seen by
transforming the definition of the smoothed cone through the
changes of variables, as we proceed to explain. Before gauge
fixing, one of the smoothed cones can be defined to behave in the
limit $\rho \to 0$ as $B_\rho=0, ~A \sim \log(\rho),~ C \sim C_0$
(while for the other we interchange $A \leftrightarrow C$). After
changing gauge and transforming to the $(X,P)$ variables this
becomes $X_1 = m/(m-1)\, \rho^{-2}, ~X_2 = const\,
\rho^{-2m/(m-1)}, ~P_1=-2/\rho, ~P_2= -2m/(m-1)\, \rho^{-2}$ which
in the $(X,\theta)$ variables tends to $(X_1,\theta_1)$ as defined
in (\ref{equilib}).


In summary, we have confirmed non-perturbatively the existence of
the smoothed cones, going beyond the perturbative analysis of
\cite{TopChange} near the smoothed tip.
Altogether there is a one-parameter family of solutions which in
the linear regime (far away from the tip) corresponds to the
various linear combinations of the two linearized modes. Two of
these solutions correspond to the smoothed cones, while all the
rest end at infinity in phase space and yield a singular geometry.
Even though strictly speaking our b.c. ``$s_+$ prescription'' is
undefined for $l=0$ since $\Re(s_1)=\Re(s_2)$, we define it to
mean to retain only the smoothed cone solutions, thereby the
double-cone is an attractor at co-dim 1, as wanted, rather than at
co-dim 2. Equivalently, these two special directions are defined
up to multiplication by $\IR_+$, and hence we loosely refer to
them as a single mode which is normally defined up to
multiplication by $\IR$.

\presub {\bf Acknowledgements}

We would like to thank Ofer Aharony for a discussion. BK
appreciates the hospitality of Amihay Hanany in MIT and KITP Santa
Barbara where parts of this work were performed.

This research is supported in part by The Israel Science
Foundation grant no 607/05 and by The Binational Science
Foundation BSF-2004117.

\appendix
\section{A numerical search for a DSS solution}
\label{search-section}


In this section we present the details of our frustrated search
for a Discretely Self-Similar (DSS) solution to the system, which
if found, would have been a candidate to be the critical merger
solution.

\presub {\bf Formulation of the problem}. The Lagrangian for the
metric in the following form
\begin{equation}
ds^{2}=e^{2B_{\rho }(\rho,\chi)}d\rho ^{2}+e^{2B_{\chi
}(\rho,\chi)}\bigl(d\chi -f(\rho,\chi)d\rho
\bigr)^{2}+e^{2\Phi(\rho,\chi)}dt^{2}+e^{2c(\rho,\chi)}d\Omega
_{D-3}^{2} ~, \label{ansatz}
\end{equation}
 where $\rho$ here is the $\log$ of $\rho$ in (\ref{def-fields}),
and choosing the gauge \bea
 f(\rho,\chi)&=&0 \non
  B_{\rho}(\rho,\chi) &=& B_{\chi }(\rho,\chi)\equiv B(\rho,\chi) \label{gauge-fix} \eea
 is
\begin{multline}
L=-e^\Psi\bigl[2B'\Psi'+(D-3)C'\bigl(\Psi'+\Phi'-C'\bigr)+({}'\rightarrow\dot{})\bigr]-\\
 (D-3)(D-4)e^{\Psi+2B-2C},
\end{multline}
where $\Psi=\Phi+(D-3)C$ as in (\ref{psidef}) and prime and dot
denote derivatives w.r.t $\rho$ and $\chi$. The constraints
obtained from the gauge fixing (\ref{gauge-fix}) are
\begin{multline}
\dot{\Phi}'+(D-3)\dot{C}'+\Phi'\dot{\Phi}+(D-3)C'\dot{C}-(B'\dot{\Phi}+\dot{B}\Phi')-(D-3)(B'\dot{C}+\dot{B}C')=0,
\end{multline}
\begin{multline}
\Phi''-\ddot{\Phi}+(D-3)(C''-\ddot{C})+\Phi'^2-\dot{\Phi}^2+(D-3)(C'^2-\dot{C}^2)-\\-2(B'\Phi'-\dot{B}\dot{\Phi})-2(D-3)(B'C'-\dot{B}\dot{C})=0,
\end{multline}
and the equations of motion which follow from the Lagrangian are
\bea
 \nabla^2 B &=& -\frac{(D-3)(D-4)}{2}e^{2b-2c}+\frac{(D-3)(D-4)}{2}\bigl(C'^2+\Dot
 C^2\bigr)+(D-3)(C'\Phi'+\Dot C\Dot\Phi) \non
 \nabla^2\Phi &=& -(D-3)(C'\Phi'+\Dot C\Dot\Phi)-\bigl(\Phi'^2+\Dot\Phi^2\bigr)\non
 \nabla^2 C &=& (D-4)e^{2b-2c}-(D-3)(C'^2+\Dot C^2)-(C'\Phi'+\Dot C\Dot\Phi)
\eea
 where $\nabla^2$ is a 2-dimensional Laplacian
$\nabla^2=\partial_{\rho\rho}+\partial_{\chi\chi}$.

We are looking for a Discretely Self-Similar (DSS) solution to
these equations, one which satisfies
$g_{\mu\nu}(\rho+\Delta_\rho,\chi)=e^{2\Delta}g_{\mu\nu}(\rho,\chi)$
for some $\Delta$ and $\Delta_{\rho}$. In our equations we change
$B\rightarrow B+\frac{\Delta}{\Delta_{\rho}}\rho$,
$\Phi\rightarrow\Phi+\frac{\Delta}{\Delta_{\rho}}\rho$,
$C\rightarrow C+\frac{\Delta}{\Delta_{\rho}}\rho$. The fields thus
defined are periodic with period $\Delta_{\rho}$. Finally we make
one more redefinition $\Phi\rightarrow\Phi+\ln{\cos{\chi/2}}$,
influenced by the form of the double-cone solution (\ref{d-cone}).
With the fields thus defined the double cone solution is constant
\bea
 B(\rho,\chi) &=& \Phi(\rho,\chi)=-\frac12\ln{\(4(D-2)\)} \non
 C(\rho,\chi) &=& \frac12\ln{\frac{D-4}{D-2}}~. \eea

The boundary conditions which the fields should satisfy are
\begin{enumerate}
 \item Reflection symmetry at the $\chi=0$ axis  \be
 \dot B(\rho,0)=\dot{\Phi}(\rho,0)=\dot C(\rho,0)=0 \label{bc1} \ee
 \item Regularity at the horizon which is at $\chi=\pi$.
There are 5 regularity conditions (3 from EOMs and 2 from
constraints), but only 4 of them are independent. In terms of the
newly defined fields these conditions read \bea
 0 = \dot B(\rho,\pi) &=& \dot{\Phi}(\rho,\pi)=\dot C(\rho,\pi)~, \non
 B'(\rho,\pi) &=& \Phi'(\rho,\pi) ~. \label{bc2} \eea
 \item Periodicity in the $\rho$ direction as mentioned above \be
  0 = \Phi(\rho+\Delta_\rho) - \Phi(\rho) = B(\rho+\Delta_\rho) - B(\rho) = C(\rho+\Delta_\rho) - C(\rho)
   \label{bc3} \ee
\end{enumerate}
Since our equations are invariant under the symmetry of adding 2
independent constants to the fields: $B\rightarrow B+const_1$,
$C\rightarrow C+const_1$, $\Phi\rightarrow\Phi+const_2$, the last
regularity condition can be integrated to the form
$B(\rho,\pi)=\Phi(\rho,\pi)$ where setting the integration
constant preserves only one remaining symmetry (scaling)
\begin{equation}
B\rightarrow B+const, C\rightarrow C+const,
\Phi\rightarrow\Phi+const.\label{freedom}
\end{equation}

\sbsection{Algorithm}. One way to find a DSS solution numerically
is the following. Using the 5 equations that we have (3 elliptic
and 2 hyperbolic) written as \be
f_i\bigl(B(\rho,\chi),\Phi(\rho,\chi),C(\rho,\chi),\Delta/\Delta_\rho
\bigr)=0 \ee
 with $i=1..5$, one constructs a ``functional''
\begin{equation}
F_0[B(\rho,\chi),\Phi(\rho,\chi),C(\rho,\chi),\Delta,\Delta_\rho]
:= \sum\limits_{i=1}^5\int\limits_0^{\pi}
d\chi\int\limits_0^{\Delta_{\rho}}d\rho
f_i^2\bigl(B(\rho,\chi),\Phi(\rho,\chi),C(\rho,\chi)\bigr).
\end{equation}
 Note that in addition to the 3 (local) fields, the functional
depends also on 2 global variables $\Delta$ and $\Delta_\rho$.
This functional vanishes if it is evaluated on a solution and
otherwise it is positive, so one can minimize it numerically. We
are interested in a solution that satisfies our boundary
conditions. Those in the $\chi$ direction (\ref{bc1},\ref{bc2})
can be incorporated in the algorithm by defining one more
non-negative functional:
\begin{multline}
F_1[B(\rho,\chi),\Phi(\rho,\chi),C(\rho,\chi)] :=
\int\limits_0^{\Delta_{\rho}}d\rho\Bigl[\dot
B^2(\rho,0)+\dot{\Phi}^2(\rho,0)+\dot
C^2(\rho,0)\Bigr]+\\+\int\limits_0^{\Delta_{\rho}}d\rho\Bigl[ \dot
B^2(\rho,\pi)+\dot{\Phi}^2(\rho,\pi)+\dot
C^2(\rho,\pi)+(B(\rho,\pi)-\Phi(\rho,\pi))^2\Bigr].
\end{multline}
The boundary conditions in $\rho$ direction (\ref{bc3}) is
incorporated differently, as will be explained below.

The remaining symmetry (\ref{freedom}) in the equations can be
fixed in many possible ways; the most symmetric one being to set
the overall average of the three fields to 0. It is done by the
following functional:
\begin{equation}
F_2[B(\rho,\chi),\Phi(\rho,\chi),C(\rho,\chi)] :=
\Bigl[\int\limits_0^{\pi}d\chi\int\limits_0^{\Delta_{\rho}}d\rho\Bigl(B(\rho,\chi)+\Phi(\rho,\chi)+
C(\rho,\chi)\Bigr)\Bigr]^2.
\end{equation}
 The total functional to be minimized is
\begin{equation}
F := F_0+F_1+F_2.
\end{equation}

In order to prevent the procedure from converging to the double
cone which is characterized by constant fields, we multiply $F$ by
$\frac12(v_{\rho}+\frac{1}{v_{\rho}})$ where the
 ``variability'' $v_{\rho}$ is defined by
\begin{equation}
v_{\rho} :=
\int\limits_0^{\pi}d\chi\int\limits_0^{\Delta_{\rho}}d\rho\Bigl(B'^2(\rho,\chi)+\Phi'^2(\rho,\chi)+
C'^2(\rho,\chi)\Bigr).
\end{equation}
The $v_{\rho}$ term is added to prevent an endless drift towards
an opposite limit, that of rapidly varying fields (since if
$v_{\rho}\rightarrow\infty$ then $\frac{F}{v_{\rho}}$ would go to
0 no matter what happens to $F$) and a factor $\frac12$ is added
in order to fix the normalization since the value of the function
$x+1/x$ at its minimum is 2.

The arbitrariness $\rho\rightarrow\rho+const$ is harmless, as will
be explained below.

\sbsection{Discretization}. By discretizing the domain $F$ becomes
a function of a large number of (local) variables:
$F=F(B_{ij},\Phi_{ij},C_{ij})$ where $B_{ij},\Phi_{ij},C_{ij}$ are
the values of fields at the grid sites. We make the grid cells
``non-square'' as follows: if the step-size in $\chi$ direction is
$h$ then we take the step-size in $\rho$ direction to be $h\cdot
d$ with $d$ being adjustable parameter (in fact if the number of
grid sites in both directions are equal then
$d=\frac{\Delta_{\rho}}{\pi+h}$). In order to impose the boundary
condition in the $\rho$ direction we calculate derivatives w.r.t.
$\rho$ at the boundaries of the grid with values of fields at the
opposite side of the grid, thus making our domain effectively
cylindrical.

One can start with some configuration of fields and change them
together with the global variables $\Delta$ and $\Delta_{\rho}$ so
that $F$ is minimized (remembering that only $F=0$ minima are
solutions). This is done by calculating the gradient $\partial
F/\partial B_{ij}$, $\partial F/\partial \Phi_{ij}$, $\partial
F/\partial C_{ij}$, $\partial F/\partial \Delta$, $\partial
F/\partial \Delta_{\rho}$ (the direction in a ``space of
parameters'' in which $F$ grows in the fastest way) and going a
bit in the opposite direction. This procedure is repeated until we
arrive at a minimum. There is an additional parameter $\Omega$
which sets the ``speed of relaxation'' by defining the change in
any parameter $a$ to be $a\rightarrow a-\Omega\, \frac{\partial
F}{\partial a}$. Instead of working with the global variables
$\Delta$ and $\Delta_{\rho}$ we find it convenient to use the
equivalent variables $d$ and
$\delta=\frac{\Delta}{\Delta_{\rho}}$. Finally, the symmetry
$\rho\rightarrow\rho + const$ is of no importance because nothing
in the algorithm produces running in this direction. Moreover,
discretization makes this symmetry discrete and creates a finite
distance between nearby degenerate vacua.

\sbsection{Results}. We ran a program with $D=5$, and so far the
only solution found is the double cone, even after using various
``tricks'' designed to avoid it, as described below. If
multiplication by $\frac12(v_{\rho}+\frac{1}{v_{\rho}})$ is not
used the program converges to the double cone very fast no matter
what the initial configuration is. If this factor is included then
it still converges to this solution with $d\rightarrow\infty$. If
one adds to $F$ an additional factor $\frac12(d+\frac{1}{d})$ then
it still converges to the double cone but in a much longer time,
so that $F$ tends to 0 faster than $1/v_{\rho}$ tends to infinity
(the configuration becomes very sensitive to changes, so it is
necessary to take $\Omega$ very small, for a grid 10$\times$10 it
should be of order $10^{-6}$ whereas without this additional
factor it works well with $\Omega\approx10^{-3}$). In all cases
$\delta$ becomes close to that of the double cone solution
$\frac{1}{2\sqrt{3}} \approx 0.2887$.

\end{document}